\documentclass[a4paper]{jpconf}

\usepackage{graphicx}
\usepackage{amsmath,amssymb,latexsym}
\usepackage{cite}
\usepackage{color}

\usepackage{epstopdf}

\def\Jl#1#2{{\it #1} {\bf #2}\ }
\def\CQG#1 {\Jl{Class. Quantum Grav.}{#1}}
\def\GC#1 {\Jl{Grav. Cosmol.}{#1}}
\def\GRG#1 {\Jl{Gen. Rel. Grav.}{#1}}
\def\PRD#1 {\Jl{Phys. Rev. D}{#1}}
\def\PRL#1 {\Jl{Phys. Rev. Lett.}{#1}}

\def\nqq{\hspace*{-2em}}

\def\noi{\noindent}

\def\d{\partial}

\def\lal{&&\nqq {}}

\def\beq{\begin{equation}}
\def\eeq{\end{equation}}
\def\bear{\begin{eqnarray}}
\def\bearr{\begin{eqnarray} \lal}
\def\ear{\end{eqnarray}}
\def\earn{\nonumber \end{eqnarray}}

\def\nnn{\nonumber\\ \lal }

\def\yyy{\\[5pt] \lal }

\def\then{\ \Rightarrow\ }

\def\half{{\tfrac{1}{2}}}

\def\qua{\tfrac 14}

\def\const{{\rm const}}
\def\eps{\varepsilon}

\def\kappa{\varkappa}
\def\e{{\rm e}}

\def\wt{\widetilde}

\def\tT{{\wt T}}

\def\S{{\mathbb S}}
\def\R{{\mathbb R}}
\def\M{{\mathbb M}}
\def\T{{\mathbb T}}

\def\ua{{\underline a}}      
\def\um{{\underline m}}
\def\ok{{\overline k}}

\def\eq{equation}
\def\eqs{equations}

\def\rf{\eqref}
\def\eqn{\eq\ \eqref}

\def\asflat{asymptotically flat}

\def\dS{de Sitter}

\def\GR{general relativity}
\def\bh{black hole}
\def\wh{wormhole}

\def\sph{spherically symmetric}
\def\ssph{static, spherically symmetric}
\def\EE{Einstein equation}

\def\fin{{\rm fin}}

\begin{document}

\title{Wormholes and black universes communicated with extra dimensions}

\author{K A Bronnikov$^{1,2,3}$, P A Korolyov$^1$, A Makhmudov\\ and M V Skvortsova$^1$} 

\address{$^1$ Peoples' Friendship University of Russia (RUDN University), 
               6 Miklukho-Maklaya St., Moscow 117198, Russia}
\address{$^2$ Center for Gravitation and Fundamental Metrology, VNIIMS,
               46 Ozyornaya St., Moscow 119361, Russia}
\address{$^3$ National Research Nuclear University MEPhI
            (Moscow Engineering Physics Institute), Kashirskoe highway 31, Moscow, 115409, Russia}
	
\ead{kb20@yandex.ru, korolyov.pavel@gmail.com, arslan.biz@gmail.com, milenas577@mail.ru}        

\begin{abstract}
       In 6D general relativity with a phantom scalar field as a source of gravity, we present solutions 
       that implement a transition from an effective 4D geometry times small extra dimensions 
       to an effectively 6D space-time where the physical laws are different from ours. 
       We consider manifolds with the structure $\M_0 \times \M_1 \times \M_2$, where $\M_0$ 
       is 2D Lorentzian space-time while each of $\M_{1,2}$ can be a 2-sphere or a 2-torus.  
       Some solutions describe wormholes with spherical symmetry in our space-time 
      and toroidal extra dimensions. Others are of black universe type: at one end there is
      a 6D asymptotically anti-de Sitter black hole while beyond the horizon the geometry tends to
      a 4D de Sitter cosmology times a small 2D spherical extra space.
\end{abstract}      	 

\section{Introduction}

  Multidimensional theories suggest a great variety of geometries,  topologies and compactification
  schemes (for reviews see, e.g.,  [1--4] and references therein). One of the opportunities of interest 
  is that the geometry can be effectively four-dimensional in some space-time region but have 
  a greater dimension in other regions. We here try to obtain examples of such space-times 
  in the framework of 6D \GR\ with a minimally coupled scalar field as a source of gravity.

  More specifically, we consider a 6D space-time of the form $\M = \M_0 \times \M_1 \times \M_2$,
  where $\M_0$ is 2D Lorentzian space-time, while each of $\M_{1,2}$ is a two-sphere or 
  a two-torus. As $x \to -\infty$, we assume an \asflat\ or asymptotically de Sitter 4D geometry 
  times small extra dimensions, while at the other end, $x\to +\infty$, we expect a geometry with 
  large extra dimensions.

  We give three explicit examples of such solutions. In two of them the 4D subspace has a 
  \wh\ geometry,\footnote
	{See, e.g., \cite{BR, visser-book, lobo} for reviews and also more recent 
                papers, e.g.,  \cite{dz1, dz2, almaz1, almaz2}.}   
  while the third example corresponds to the notion of a black universe, that is, a \bh\ with 
  an expanding universe instead of a singularity beyond its horizon \cite{bu1, bu2, bu3}.   

  In all these cases, the solutions can only be obtained with a phantom scalar field, 
  having a wrong sign of kinetic energy. Such fields naturally emerge in some unification theories,    
  there are theoretical arguments both {\it pro et contra} their possible existence, see, e.g., 
  discussions in \cite{bsta1, BFab07, visser-book, lobo}. In this paper, as in many 
  others, we admit it as a working hypothesis.
 
\section{Equations in 4+2 dimensions}

  We consider 6D GR with a minimally coupled scalar field $\phi$ having a potential $V(\phi)$
  as the only source of gravity. The total action is 
\beq         \label{act}
             S = \frac{m_6^2}{2} \int \sqrt{|g_6|}
	 \Big[R_6 + 2\eps_\phi g^{AB} \d_A\phi \d_B\phi - 2V(\phi)\Big],
\eeq
  where $m_6$ is the 6D Planck mass, $R_6$ and $g_6$ are the 6D Ricci scalar and metric
  determinant, respectively, $\eps_\phi$ is $+1$ for a normal, canonical scalar field and $-1$ for 
  a phantom one, and $A, B, \ldots = \overline{0,5}$.  The corresponding equations 
  of motion are the scalar field equation $2\eps_\phi \Box_6\phi + dV/d\phi =0$ and the 
  \EE s which can be written as
\bearr             \label{EE}
             R^A_B = - \tT^A_B \equiv - T^A_B - \qua \delta^A_B T^C_C 
                         \equiv - 2\eps_\phi  \d^A\phi \d_B\phi + \half V(\phi) \delta^A_B,
\ear
  where $R^A_B$ is the 6D Ricci tensor and $T^A_B$ is the stress-energy tensor 
  (SET) of the scalar field.

  Consider a 6D manifold being a direct product of three 2D spaces,
  $\M = \M_0 \times \M_1 \times \M_2$, where $\M_0$ is 2D space-time 
  with the coordinates $x^0 = t$ and $x^1 =x$, while $\M_1$ and 
  $\M_2$ are compact 2D spaces of constant nonnegative curvature, i.e., each of them can
  be a sphere or a torus. The metric is taken in the form:
\beq                                                 \label{ds_6}
	ds^2 = A(x) dt^2 - dx^2/A(x) - R(x) d\Omega_1^2 - P(x) d\Omega_2^2,
\eeq 
  where $A(x),\ R= r^2(x),\ P= p^2(x)$ are functions of the ``radial'' coordinate $x$, chosen 
  under the condition $g_{tt}g_{xx}=-1$ (the so-called quasiglobal gauge \cite{BR}),
  while $d\Omega_1^2$ and  $d\Omega_2^2$ are $x$-independent metrics 
  on 2D manifolds $\M_1$ and $\M_2$ of unit size. We also assume $\phi = \phi(x)$.    

  We do not fix which of $\M_{1,2}$ belongs to our 4D space-time and which is ``extra'':
  everything depends on their size. Thus, if $\M_1$ is large and spherical while $\M_2$ 
  is small and toroidal, we have a \ssph\ configuraton in 4D and a toroidal extra space, and so on. 

  Due to the symmetry of the problem and the properties of the scalar field SET, 
  there are four independent equations, which may be written as follows
  (the prime denotes $d/dx)$):
\bearr                                  \label{00}
             R^t_t = - \tT^t_t \ \then \ \ - (PR)^{-1}(A' PR)' = V(\phi),
\\ \lal  	                                 \label{01}     
             R^t_t {-} R^x_x = -\tT^t_t {+}\tT^x_x  \  \then \ \
				\frac{r''}{r} + \frac{p''}{p} = -\eps_\phi  \phi'{}^2,
\\ \lal                                  \label{02}
             R^t_t {-} R^{\ua}_a = 0  \ \then \ \ [P (AR' - A'R)]' = 2\eps_1 P, 
\yyy                                    \label{05}
             R^t_t {-} R^{\um}_m = 0 \ \then \ \ [R (AP' - A'P)]' = 2\eps_2 R,
\ear
  where $a= 2,3$ (belong to $\M_1$, $m=4,5$ (belong to $\M_2$, there is no summing over an 
  underlined index, $\eps_1 = 1$ if $\M_1$ is a sphere and $\eps_1 =0$ if it is a torus, and similarly for
  $\eps_2$ and $\M_2$.
 
  Equations \rf{02} and \rf{05} contain only the metric functions $A(x), P(x),  R(x)$. 
  Therefore, considering them separately, these are two equations for three unknown 
  functions, so there is arbitrariness in one function. If the metric functions are known, the other 
  two \EE s can be used to find the scalar $\phi$ and the potential $V$. From \eqn{01} it 
  follows that solutions with $r > 0$ and $p > 0$ in the whole range $x \in \R$ can only exist
  with $\eps_\phi = -1$, i.e., a phantom field, since such solutions require $r''>0$ and $p'' >0$. 

\section {Possible asymptotic behaviour of the metric}

  The metric under consideration describes the following types of geometries:

\medskip\noi
  (i) SS (double spherical) space-times: the case $\eps_1 = \eps_2 =1$.

\medskip\noi
 (ii) ST (spherical-toroidal) space-times: the case $\eps_1 =1,\ \eps_2 =0$ or vice versa. 
  If $M_1$ is large and $\M_2$ small, we have static spherical symmetry in our space-time 
  and a small toroidal extra space. The opposite situation is also possible as well as a total observable 
  6D geometry.

\medskip\noi
  (iii) TT (double toroidal) space-times: if $\eps_1 = \eps_2 =0$, we have the same as 
  before but both $\M_1$ and $\M_2$ are toroidal. 
  
  We seek configurations where $x \in \R$ and there are different geometries
  in the two asymptotic regions $x\to \pm\infty$. In particular, there can be a 4D flat 
  asymptotic region at large negative $x$ times small extra dimensions and something different 
  at the other end. In this section we do not consider the properties of the scalar field but only 
  analyze which kinds of asymptotic behaviour are compatible with \eqs \rf{02} and \rf{05} 
  for each of  the types 1--3 of 6D geometry.  
  
  As an example, let us consider an \asflat\ 4D space-time with constant extra dimensions
  in SS geometry.  Without loss of generality, this means that
\beq                                                                                                      \label{flat+}
	A(x) \to \fin, \quad\ R(x) \sim x^2, \quad\ P(x) \to \fin \quad {\rm as}\ \ x\to \infty.
\eeq
  Let us substitute these conditions to \eqs \rf{02} and \rf{05}. According to \rf{flat+},
  $R' \sim x$, $A' \sim x^{-2}$ or even smaller (due to the expansion 
  $A = A_- + A_{-1}/x + \ldots$), and the l.h.s. of \rf{02} tends, in general, to a nonzero 
  constant, which agrees with the requirement to $P$ that stands on the r.h.s.. However, in 
  \rf{05} the expression in square brackets tends to a constant, hence its derivative 
  vanishes, while the r.h.s., equal to $2R$, should behave as $x^2$. We conclude that 
 {\it the asymptotic conditions \rf{flat+} are incompatible with the field equations.}

  On equal grounds we could consider $x\to -\infty$ and/or exchange $R(x)$ and $P(x)$. 
  Other opportunities are considered in the same manner, and the results are summarized in 
  table 1.  

\begin{table}[t]
\caption{\small Asymptotic behaviours compatible with \eqs\ \rf{02} and \rf{05}. Notations: a 
           plus or minus mean that a particular behaviour is possible or impossible, 
           respectively, $\pm$ that it is possible under special conditions for the functions involved. 
           The symbol ``$\fin$'' means a positive constant, while an asterisk indicates that this asymptotic 
	  is only possible with $A<0$.  }
\medskip
\centering
\begin{tabular}{cccccllll}
\hline\\[-9pt]
         &\multicolumn{3}{c}{Asymptotic behaviour }&& \multicolumn{3}{c}{6D geometries\ \ \ }&  \\
                                                          \cline{2-4}\cline{6-8}
   \raisebox{6pt}{Line No.} & $A(x)$ &\ $R(x)$ & $P(x)$   &&     SS   &   ST   & TT   &  
                                         \raisebox{6pt}{Comments}   \\ 
\hline\\[-6pt]
   0& $\fin$ &\ $\fin$ & $\fin$     &&   --        &   \ --     & $\pm$ & $\M^2\times \T^2\times \T^2$\\
   1& $\fin$ &\ $\fin$ & $x^2$     &&   --        &  \ --     &  --  & none\\ 
   2& $\fin$ &\ $x^2$ & $\fin$     &&   --        &  \ +      &  --  &  $\M^4 \times \T^2$ \\ 
   3& $\fin$ &\ $x^2$ & $x^2$     &&   +        & \  --      &  --  &   $\M^6$ \\
   4& $x^2$ &\ $\fin$ & $\fin$     &&   $+*$   & \ --       &  --  &   dS$_2\times \S^2\times \S^2$\\
   5& $x^2$ &\ $\fin$ & $x^2$     &&$\pm*$  & \ $\pm*$& -- &   dS$_4\times \S^2$\\ 
   6& $x^2$ &\ $x^2$ & $\fin$     &&$\pm*$  & \  --       &  --  &  dS$_4\times \S^2$\\   
   7& $x^2$ &\ $x^2$ & $x^2$     &&  $\pm$  &\ $\pm$  &$\pm$ &  (A)dS$_6$    \\[5pt] 
\hline
\end{tabular}
\end{table}
   
  The table shows that the choice of possible 6D geometries is rather restricted.       
  For example, SS space-times cannot have a flat Minkowski asymptotic times a finite sphere,
  but a \dS\ behaviour times a finite sphere can be realized. In ST geometry, in addition to 
  effectively 6D asymptotics, there can exist an \asflat\ \sph\ 4D space-time with constant 
  extra dimensions. Evidently, at the two ends, $x \to \pm\infty$, one can have similar
  or different admissible asymptotics. In what follows we will give three examples of 
  such solutions to the field equations where on one end there is a 4D geometry with small 
  extra dimensions and on the other there is an effectively 6D space-time.

\section {Examples}

\subsection*{Example 1: ST geometry, wormholes with a massless scalar}  

  Of  interest are space-times with \asflat\ \sph\ geometry in one asymptotic region 
  and something different in the other. The first example is an ST wormhole geometry with 
  a strongly different size of extra dimensions at the two ends, which exists among well-known 
  solutions for a massless scalar, $V=0$ \cite{k95, bim97}. One of such solutions has the form
  \cite{bs16} (note that the coordinate $z$ is different from $x$ used in the rest of the paper) 
\bearr                                       \label{ds_2}
       ds^2 = dt^2 - \e^{-4nu}\big[dz^2 + (z^2+ k^2) d\Omega_1^2\big] -\e^{2nu}d\Omega_2^2,
\nnn
         \phi = Cu \equiv (C/k)\cot^{-1} (-z/\ok),	
\ear
  where $n > 0,\ C,\ k>0$ are integration constants related by $2C^2 = k^2 + 3 n^2$. 
  It describes a \sph, twice \asflat\ wormhole in the 4D subspace $\M_0 \times \M_1$
  with a toroidal extra space $\M_2$ having a unit size, $p_-$ at $u=0$ 
  (that is, $z = -\infty$) and the size $p_+ = \e^{n\pi/k} p_-$ at the other end, 
  $u = \pi/k$, corresponding to $z = + \infty$.\footnote
	{Both ends correspond to line 2 in table 1. In the trivial case $n=0$ we obtain the 
            well-known 4D Ellis \wh\ \cite{k73, hell} times a toroidal extra space of constant size.}

  Suppose that the size of extra dimensions $p_-$ on the left end, $z=-\infty$, is small 
  enough to be invisible by modern instruments, say, $p_- = 10^{-17}$ cm.  
  The size $p_+$ on the other end is much larger if $n/k$ is large enough.
  For example, to obtain $p_+ \sim 1$ m, one should take $n/k \approx 14$.

  The \wh\ throat is a minimum of $g_{22} = \e^{-4nu}(z^2+ k^2)$, its radius is equal to
\beq               \label{r_th} 
	r_{\min} = \sqrt{k^2 + 4n^2} \exp{\Big(\frac{2n}{k} \cot^{-1} \frac{2n}{k}\Big)}.
\eeq   
  To make this radius large enough for passing of a macroscopic body, large values of $k$
  are required: e.g., to obtain $r_{\min} = 10$ m, one has to suppose $k \sim 10^{18}$.

\subsection*{Example 2: ST geometry, asymptotically AdS wormholes}

  With nonzero potentials $V(\phi)$, in most cases solutions can be found 
  only numerically. An exception appears if after integrating \eqn{05} we turn the appearing 
  constant to zero. It then follows $P = cA$, $c = \const $, and \eqn{02} takes the form
  $A^3 (R/A)']' = 2A$. It is a single equation for two functions $A(x)$ and $R(x)$, which can 
  be solved by quadratures if one specifies $A(x)$ since it can be rewritten as
\beq                      \label{R'}
	      \Big(\frac RA\Big)' = \frac {2}{A^3} \int A(x) dx.
\eeq   
  To obtain an example with an \asflat\ 4D space-time on the left end and an AdS asymptotic 
  on the right (a transition from line 2 to line 7 in table 1), 
  we should suppose $A \to 1$ as $x\to -\infty$ and $A \sim x^2$ as $x \to +\infty$. 
  It is  hard to find $A(x)$ with such properties that would lead to good analytic expressions 
  of other quantities. Therefore, an example has been obtained 
  with a piecewise smooth function $A(x)$ \cite{bs16}:
\beq             \label{A(x)}    
               A(x) = 1,\ \  x \leq 0;\qquad A(x) = 1 + 3 x^2/a^2, \ \ x\geq 0 \qquad  (a = \const > 0). 
\eeq
  The resulting solution has a $\rm C^1$ smooth metric but jumps in $\phi'$ and $V$ (figure 1).
  We have $P = cA$ in the whole space and the following behavior of $R(x)$: 
 \beq                                               \label {R-}
             R(x) = x^2 + b^2 \ \  (x \leq 0),\qquad
             R(x) = \Big(1 + \frac{3 x^2}{a^2}\Big) \biggl[b^2 
                                    + \frac{x^2 (1 + 2 x^2/a^2)}{(1 + 3 x^2/a^2)^2}\biggr] \ \  (x \leq 0),
\eeq  
  with $b = \const > 0$ (thus $x=0$ is a throat of radius $b$). At negative $x$ we have 
  $V(x) \equiv 0$ and $\phi(x) = \arctan(x/b)$ while at positive $x$ the expressions are 
  rather cumbersome.
\begin{figure*}[h]
\centering
\includegraphics[width=6.5cm,height=4.3cm]{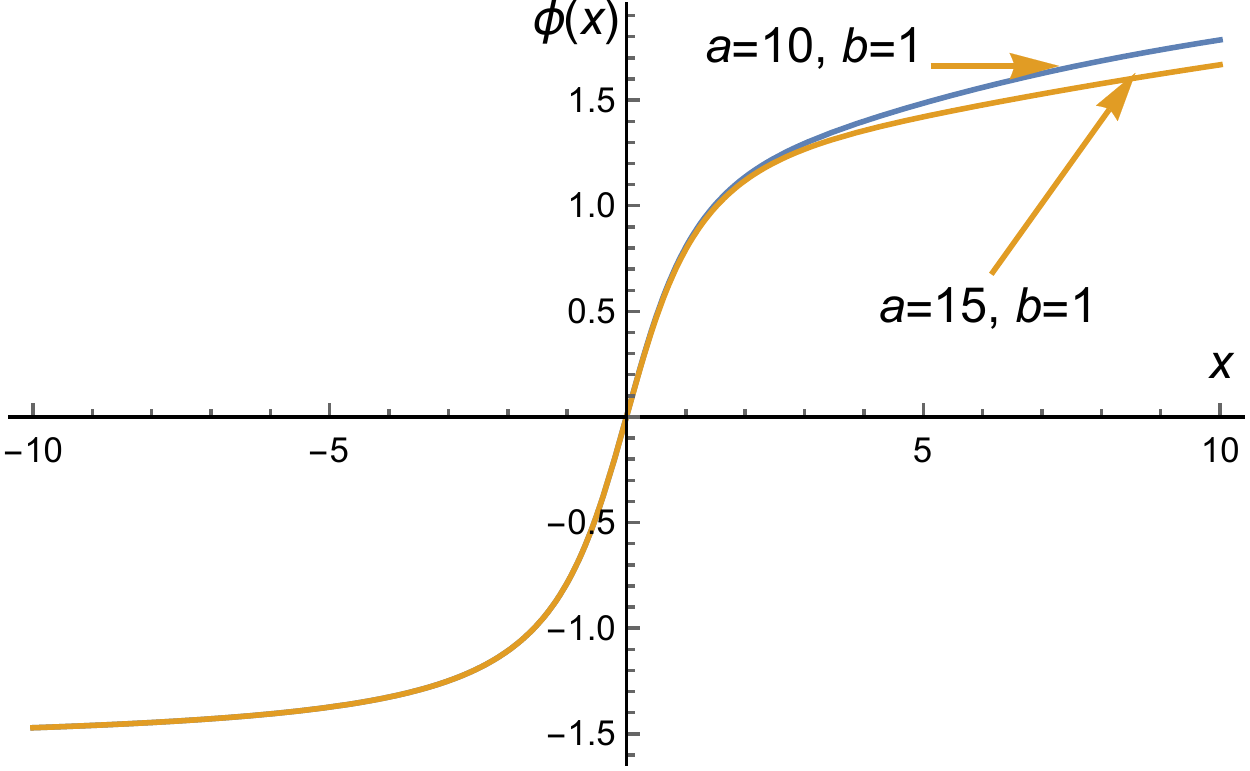}
\qquad
\includegraphics[width=6.5cm,height=4.3cm]{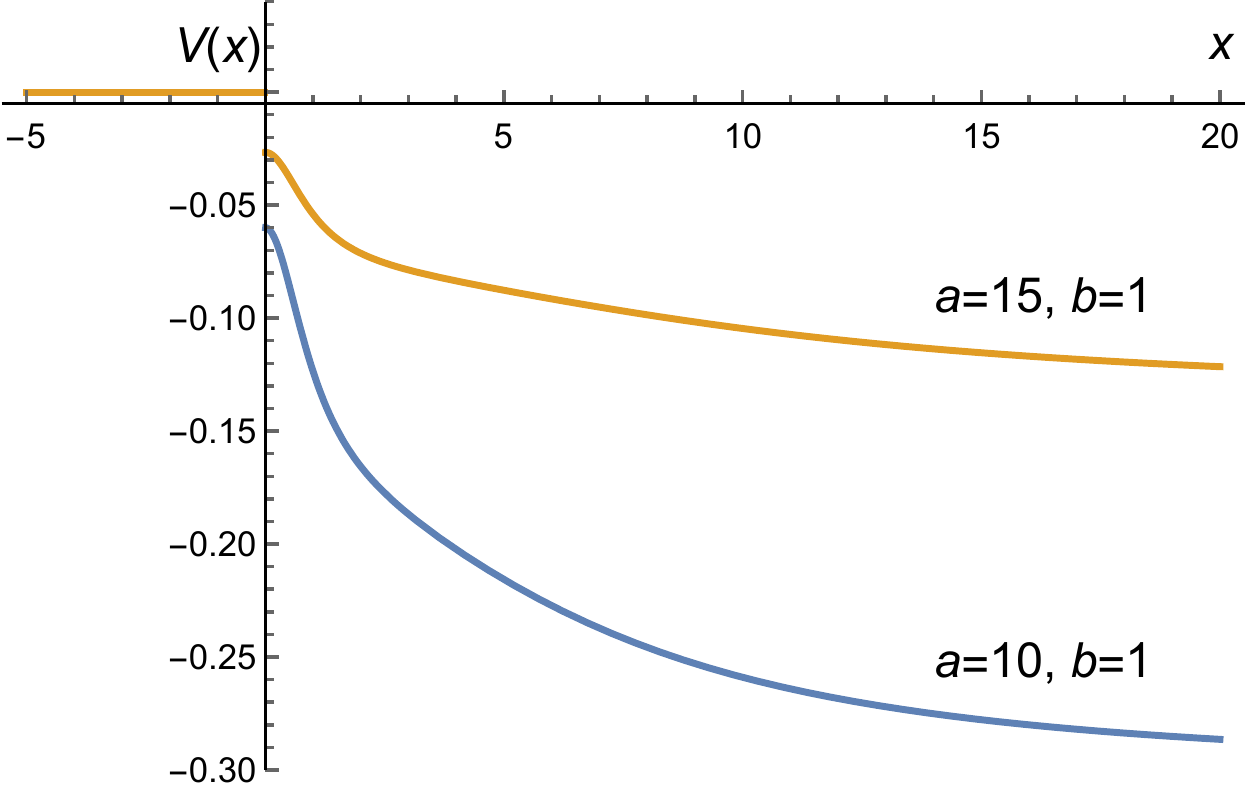}
\caption{\small The scalar field $\phi(x)$ (left) and the potential $V(x)$ (right) in example 2}
\end{figure*}
  
  The resulting configuration is \asflat\ times constant (arbitrarily small) extra dimensions 
  on the left end and a 6D AdS asymptotic on the right end, with $V$ tending to a negative constant   
  playing the role of a cosmological constant.  

  The jumps in $V(x)$ and $\phi'(x)$ at $x=0$ could be easily removed by choosing $A(x)$ 
  smoother than $\rm C^1$ at $x=0$, which is possible by making a suitable arbitrarily small 
  addition to \rf{A(x)}. 

\subsection*{Example 3: SS geometry, a black universe with $\rm AdS_6$ at the far end}

  In SS geometry, corresponding to $\eps_1=\eps_2 =1$, it is hard to solve equations \rf{02} 
  and \rf{05} analytically, but examples of interest can be obtained numerically. One such solution, 
  which has been found under the assumption $R = 1+x^2$, realizes a transition from line 6 to line 7 
  of table 1. The metric functions are plotted in figure 2, the corresponding $\phi(x)$ and $V(x)$ 
  are not shown but are easily obtained using \rf{00} and \rf{01}. On the left end ($x\to -\infty$)
  there is a 4D de Sitter metric (describing an exponentially expanding or contracting spherically 
  symmetric space) times small extra dimensions; on the right end there is an AdS static space-time.
  The configuration as a whole is a black universe \cite{bu1, bu2} where the cosmological 
  expansions starts from a horizon, the static region belonging to the remote past; in this case this
  expanding universe has a multidimensional origin.   

\begin{figure*}[h]
\centering
\includegraphics[width=7.5cm,height=5cm]{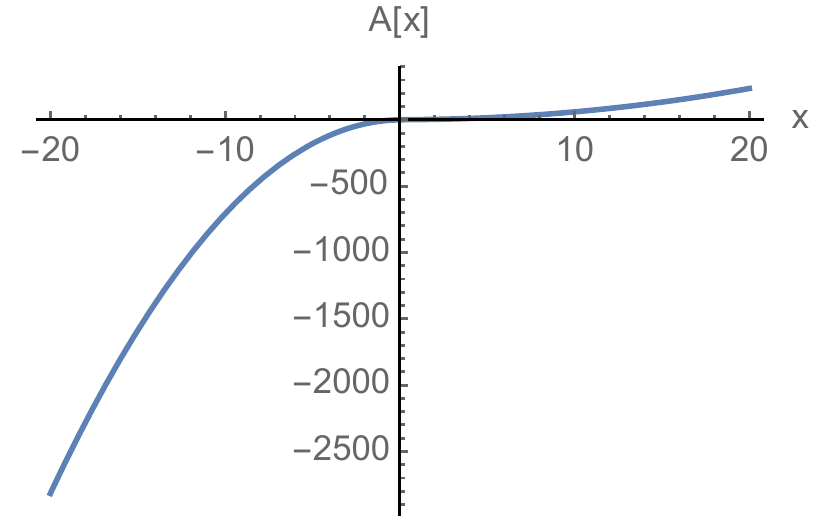}
\qquad
\includegraphics[width=7.5cm,height=5cm]{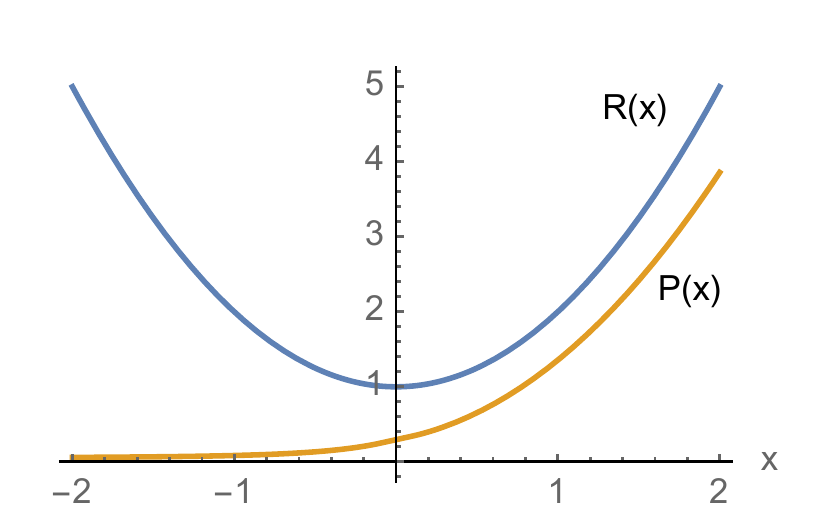}
\caption{\small The metric functions in example 3. Left: $A(x)$, with the asymptotics
    $A(-\infty) \approx -7.02 x^2$  and $A(+\infty) \approx 0.59 x^2$.
    Right: $R = 1+x^2$, $P \to 0.0474$ as $x\to -\infty$ and $P(+\infty)\approx 0.8 x^2$.}
\end{figure*}

\section{Concluding remarks}

  The results can be summarized as follows. In 6D GR with a minimally coupled scalar field as 
  a source of gravity, we have constructed examples of space-times which are effectively 4D
  on the left end and effectively 6D on the right end. Some of them have the nature of wormholes,
  others (black universes) actually represent expanding cosmological models of multidimensional
  origin.  The existence of such configurations or their analogs with a different number of extra
  dimensions in our universe cannot be a priori excluded, and their possible astrophysical 
  consequences could be a subject of further studies.   

  It should be noted that our analysis certainly did not cover all opportunities: other, more complicated 
  cases are also possible. For instance, of particular, there can be models where different 
  subspaces exchange their roles as those belonging to a large observable space, like that 
  described in \cite{Rub16}. One more subject of a future study can be a relationship between 
  the present scalar-vacuum system and multidimensional gravity with curvature-nonlinear 
  actions \cite{BR, Rub16} in different conformal frames in application to space-times of the 
  types considered here and in \cite{Rub16}.  

\subsection*{Acknowledgments} 

 We thank Sergei Rubin and Sergei Bolokhov for numerous helpful discussions. 
 The work of KB was partly performed within the framework of the Center 
  FRPP supported by MEPhI Academic Excellence Project 
  (contract No. 02.03.21.0005, 27.08.2013).
  This work was also funded by the Ministry of Education and Science of the Russian
  Federation on the program to improve the competitiveness of the RUDN University among the 
  world leading research and education centers in 2016--2020 and by RFBR grant 16-02-00602.

\def\arXiv{{\it arXiv}:\ }

\end{document}